\documentstyle[epsf]{elsart}
\begin{document}
\begin{frontmatter}
\title{Distribution of fermionic and topological observables 
       on the lattice}

\vspace{-16.5mm}
\hspace{7.5cm} \footnote{Supported in part by FWF under Contract No. P11456. } 

\vspace{10mm}

\author[]{W. Sakuler,}  
\author[]{S. Thurner and}
\author[]{H. Markum}
\address[{}]{
 Institut f\"ur Kernphysik, TU Wien, Wiedner  Hauptstra\ss{}e 8-10,
 A-1040 Vienna, Austria }

\begin{abstract}
We analyze  the topological and fermionic vacuum structure 
of four-dimensional QCD on the lattice by means of correlators of 
fermionic observables and topological densities. 
We show the existence of strong  
local correlations between the  topological charge and monopole 
density on the one side and the quark condensate, charge and chiral density
on the other side. 
Visualization  of individual gauge configurations demonstrates 
that instantons (antiinstantons)  
carry positive (negative) chirality, whereas the quark charge density 
fluctuates in sign within instantons. \\
\vspace{3mm}
%\vspace{3mm}

\end{abstract}
\end{frontmatter}

\section{Introduction and Theory}

During the last two decades several models have been developed to 
describe the basic properties of QCD, quark confinement 
and chiral symmetry breaking. The most popular are 
the dual superconductor, leading to confinement, and 
the instanton liquid model, which explains chiral symmetry breaking 
and solves the $U_A(1)$ problem \cite{SHU88}. Both  models 
rely  on the existence of topological excitations, 
monopoles and instantons.  
Instantons have integer topological charge $Q$ which is associated with  
the homotopy group $\pi_3(SU(N_c))=Z$, in four-dimensional 
Euclidian space. It is related to  
the zero eigenvalues of the massless Dirac operator $D\!\!\!\!/\,\,\,$ 
via the Atiyah-Singer index theorem \cite{singer}:
\begin{equation}
Q={\mathrm Index}(D\!\!\!\!/\,\,\,)=n_+ - n_- \,\, ,
\end{equation}
where $n_+$ $( n_-)$ is the number of zero-modes of positive (negative) 
chirality. 
Apart from this famous connection of topology and fermionic 
degrees of freedom, here we attempt to systematically shed light 
on the relationship between the sea-quark distribution and 
topological density. 
We do this by studying correlators of topological densities  
with fermionic observables  of the form 
$ \bar\psi\Gamma\psi$ with $\Gamma = 1,\gamma_4,\gamma_5$.
Those quantities are usually referred to as the  
quark condensate, quark charge density, and the chiral density. 

Recently, it was demonstrated that monopole currents which constitute 
a different topological excitation of compact SU(3) gauge theory, 
related to the homotopy group $\pi_2 (SU(N_c)/U(1)^{N_c-1} ) = Z^{N_c-1}$, 
 appear preferably 
in the regions of non-vanishing topological charge density \cite{wir,andere}.
It has been discussed  that both instantons and monopoles are related 
to chiral symmetry breaking \cite{DIA84,MIA95,warsch96}. 
In \cite{sasaki98} it was shown that zero-modes do not appear 
in configurations which are 
not monopole dominated. 
Here, by interpreting the correlators of monopole densities and  
fermionic operators, we confirm the local interrelation of monopole and 
fermionic degrees of freedom.   
%%%%%%%%%%%%%%%%%%%%%%%%%%%%%%%%%%%%%%%%%%%%%%%%%%%%%

For the implementation of the topological charge on the lattice
there exists no unique discretization. In this work we
restrict ourselves to the so-called field theoretic definitions which
approximate the topological charge in the continuum, 
\begin{equation}
q(x)=\frac{g^{2}}{32\pi^{2}} \epsilon^{\mu\nu\rho\sigma}
\ \mbox{\rm Tr} \ \Big ( F_{\mu\nu}(x) F_{\rho\sigma}(x) \Big ) \ ,
\end{equation}
in the following ways \cite{divecchia}:
\begin{equation}
 q^{(P,H)}(x)=-\frac{1}{2^{4}32\pi^{2}}
\sum_{\mu,\ldots =\pm 1}^{\pm 4}
\tilde{\epsilon}_{\mu\nu\rho\sigma} \mbox{\rm Tr} 
\ O_{\mu\nu\rho\sigma}^{(P,H)} ,
\end{equation}
with
\begin{equation}
O_{\mu\nu\rho\sigma}^{(P)} = U_{\mu\nu}(x) U_{\rho\sigma}(x) \ ,
\end{equation}
for the plaquette prescription and
\begin{eqnarray}
O_{\mu\nu\rho\sigma}^{(H)} &=&
   U(x,\mu) 
	U(x\!+\!\hat\mu,\nu) 
	U(x\!+\!\hat\mu\!+\!\hat\nu,\rho) 
   U(x\!+\!\hat\mu\!+\!\hat\nu\!+\!\hat\rho,\sigma) 
\nonumber \\ & \times &
   U^{\dagger}(x\!+\!\hat\nu\!+\!\hat\rho\!+\!\hat\sigma,\mu) 
   U^{\dagger}(x\!+\!\hat\rho\!+\!\hat\sigma,\nu)
   U^{\dagger}(x\!+\!\hat\sigma,\rho) 
	U^{\dagger}(x,\sigma), 
%\nonumber \\ 
\end{eqnarray}
for the hypercube prescription.
We mention here that  the  topological charges
employed are locally gauge invariant, whereas the monopole currents are not.
The lattice and continuum versions of the theory represent
different renormalized quantum field theories, which differ 
by finite, non-negligible renormalization factors. 
A simple procedure  to get rid of renormalization constants,
while preserving physical information contained in lattice configurations,
is the cooling method.
The cooling procedure systematically reduces quantum fluctuations, 
and suppresses
differences between the different definitions of the topological charge.
In our investigation we have employed the so-called 
``Cabibbo--Marinari method'' which consists in a local minimization of the gluonic 
action in the SU(2) subgroups of SU(3) \cite{cm}.

In order to investigate monopole currents one has to project $SU(N)$
onto its  abelian degrees of freedom, such that an abelian $U(1)^{N-1} $
theory remains \cite{thooft2}. This aim can be achieved by
various gauge fixing procedures. We employ the
so-called maximum abelian gauge which is most favorable for our
purposes.
For the definition of the monopole currents $m(x,\mu)$ we use the
standard method \cite{kronfeld87}. 
To extract abelian parallel transporters $u(x,\mu)$ 
after imposing the maximum abelian gauge one has to perform
the decomposition 
\begin{equation}
\tilde{U}(x,\mu) = c(x,\mu) u(x,\mu)  \ , 
\end{equation}

with $(N=3)$

\begin{displaymath} 
\label{ab_u}
u(x,\mu) = \mbox{\rm diag } [ u_{1}(x,\mu), u_{2}(x,\mu), u_{3}(x,\mu) ] \ , 
\end{displaymath}
\begin{displaymath} 
u_{i}(x,\mu) = \exp \Big [ i \ \mbox{\rm arg } \tilde{U}_{ii}(x,\mu) -
\frac{1}{3} i \phi(x,\mu) \Big ] \ , 
\end{displaymath}
\begin{equation} 
\phi(x,\mu) = \sum_{i} \mbox{\rm arg } \tilde{U}_{ii}(x,\mu) \Big | _{\mbox{\tiny mod $2\pi$}} \! \in \! (-\pi,\pi] \ . 
\end{equation}
Since the maximum abelian subgroup $U(1)^{N-1}$ is compact, there
exist topological excitations. These are color magnetic monopoles which have
integer-valued magnetic currents on the links of the dual lattice:
\begin{equation}
m_{i}(x,\mu) = \frac{1}{2\pi} \sum_{\Box \ni \partial f(x+\hat\mu,\mu)}
\mbox{\rm arg } u_{i}(\Box) \ ,
\end{equation}
where $u_{i}(\Box)$ denotes a product of abelian links $u_{i}(x,\mu)$ around
a plaquette $\Box$ and $f(x+\hat\mu,\mu)$ is an elementary cube perpendicular
to the $\mu$ direction with origin $x+\hat\mu$.
The magnetic currents form closed loops on the dual lattice as a consequence
of monopole current conservation.
From the monopole currents we define the local monopole density as
\begin{equation}
 \rho(x) = \frac{1}{ 3 \cdot 4} \sum_{\mu,i} | m_{i}(x,\mu) | \ .  
\end{equation}

%%%%%%%%%%%%%%%%%%%%%%%%%%%%%%%%%%%%%%%%%%%%%%%%%%%%%
%After fixing the gluon field to the maximum Abelian gauge, 
%\begin{equation}
%(\partial_{\mu} \pm igA^3_{\mu})A^{\pm}_{\mu}=0 \,\,\, , \,\,\, 
%A^{\pm}_{\mu}= A^1_{\mu} \pm iA^2_{\mu} \quad ,
%\end{equation}
%Abelian color magnetic monopole currents $m(x,\mu)$ are evaluated over elementary
%cubes in the standard way used on the lattice \cite{kronfeld87}. 
%The quantity of further interest is the monopole 
%density being defined as $\rho(x)=\frac{1}{4}\sum_{\mu}|m(x,\mu)|$.
The local quark condensate $\bar \psi \psi (x)$ 
is a diagonal element of the inverse of the fermionic matrix
of the QCD action. Other fermionic operators 
are obtained by inserting the Euclidian $\gamma_4$ 
and $\gamma_5$ matrices. 
We compute correlation functions between two observables ${\cal O}_1(x)$ and ${\cal O}_2(y)$ 
\begin{equation}
\label{correlations}
g(y-x)=\langle {\cal O}_1(x) {\cal O}_2(y) \rangle - 
       \langle {\cal O}_1\rangle \langle {\cal O}_2\rangle. 
\end{equation}
Since  topological objects with opposite sign are equally distributed,
we correlate the  
quark condensate with the square of the topological charge density,
and similarly for the other quantities.

%%%%%%%%%%%%%%%%%%%%%%%%%%%%%%%%%%%%%%%%%%%%%%%%%%%%%%%%%%%%%%%
\section{Results}
Our simulations were performed for full SU(3) QCD on an 
$8^{3} \times 4$ lattice with
(anti) periodic boundary conditions for the (fermionic) gluonic 
sector.  
Applying a standard Metropolis algorithm,  
we checked  that tunneling between sectors of different topological 
charges occurs at reasonable rates. 
Dynamical quarks in Kogut-Susskind discretization   
with $n_f=3$ flavors of degenerate  mass $ma=0.1$ were taken into account using the 
pseudofermionic method with 40 updates for the 
production runs. 
We performed runs  in the confinement phase at inverse gluon coupling 
$\beta=5.2$, corresponding to a lattice spacing $a \approx 0.2$ fm. 
This leads to a total extension of the lattice of $1.6$ fm and an energy 
unit of $1$ GeV. 
Measurements were taken on 2000 configurations separated by 50 sweeps.

%23%%%%%%%%%%%%%%%%%%%%%%%%%%%%%%%%%%%%%%%%%%%%%%%%%%%%%%%%%%%%%%
\begin{figure*}
\begin{tabular}{c}
\epsfxsize=13.5cm\epsffile{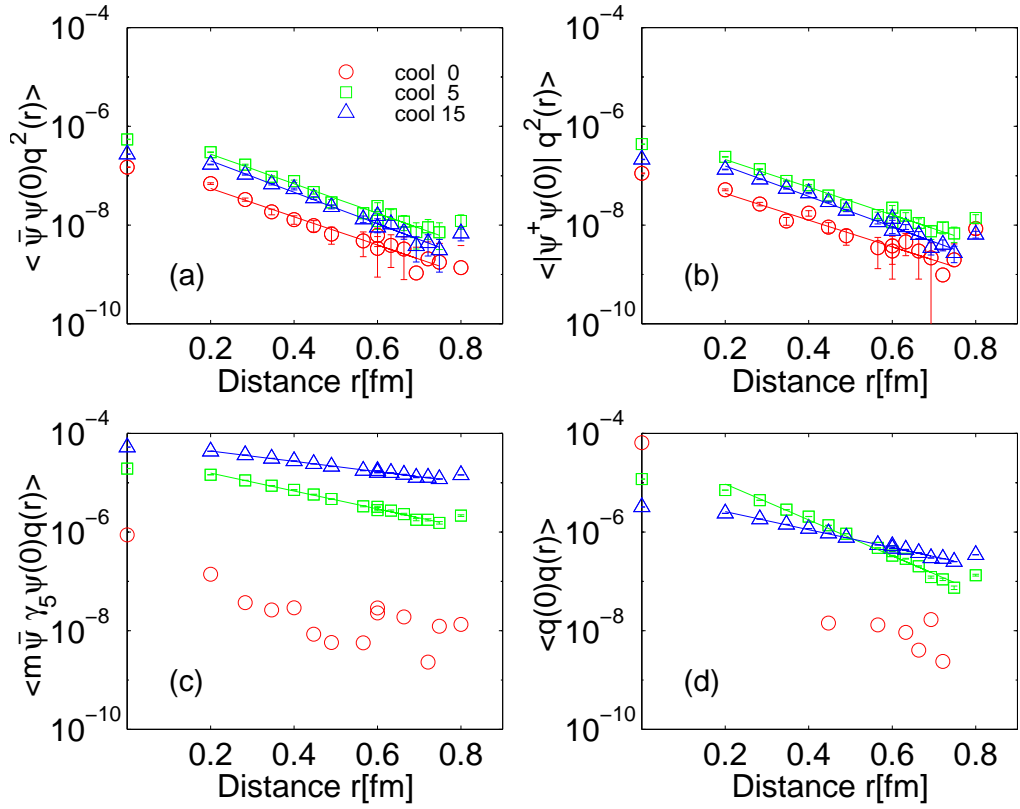} \\
\end{tabular}
\caption{
Correlation functions of the topological charge density  
with the quark condensate (a), quark charge (b) 
and chiral density (c), and its autocorrelation (d). 
For clarity of the plots error bars extending to negative values
have been omitted. 
In the plots (c) and (d) for zero cooling steps some 
points cannot be resolved or have very  
large errors.} 
\label{corr}
\end{figure*}
%%%%%%%%%%%%%%%%%%%%%%%%%%%%%%%%%%%%%%%%%%%%%%%%%%%%%%%%%%%%%%%
Figure~\ref{corr} shows results for the correlation functions 
of Eq.~(\ref{correlations}) with 
${\cal O}_1$ a local fermionic observable (except in (d)) and $ {\cal O}_2$
 the  topological charge density for different numbers of cooling sweeps
 \cite{boulder98}. 
The correlations of the two densities are given in lattice units  
which leads to the small absolute values. The screening masses 
obtained from 
fits of the correlators to an exponential are collected in 
Table 1. The fits were taken over the interval indicated by the lines in 
Fig.~\ref{corr}. They clearly hint at a linear relation in the 
logarithmic scale. 
The statistical errors of the standard deviation 
in the plots are generally small, 
leading to relatively small errors in the fit parameters. 
The extracted quantities have to be interpreted as 
effective masses and reflect the nonperturbative gluon exchange. 
Although cooling of quantum fields is necessary to extract topological 
structure, the exponential slopes of the correlators between 
the topological charge density  and 
both the local chiral condensate (a) and the absolute value of the quark 
charge density (b) are nearly cooling-independent. 
%26%%%%%%%%%%%%%%%%%%%%%%%%%%%%%%%%%%%%%%%%%%%%%%%%%%%%%%%%%%%%%%
\begin{table}[t]
\begin{center}
\begin{tabular}{l c c c }
\hline
Correlation                          & cool 0   & cool 5   & cool 15  \\
\hline
$\bar\psi\psi(0) q^2(r)$             & 1.32(34) & 1.38(16) & 1.47(16) \\
$|\psi^{\dagger}\psi(0)| q^2(r)$     & 1.25(66) & 1.29(10) & 1.42(09) \\
$\bar\psi \gamma_5\psi(0) q(r)$      &  -       & 0.84(02) & 0.48(01) \\
$q(0) q(r)$                          &  -       & 1.67(02) & 0.84(01) \\
\hline
$\bar\psi\psi(0) \rho(r)$            & 1.14(10) & 1.16(05) & 1.27(06) \\
$|\psi^{\dagger}\psi(0)| \rho(r)$    & 1.13(02) & 1.08(01) & 1.16(01) \\
$|\bar\psi\gamma_5\psi(0)|^2 \rho(r)$&  -       & 0.98(07) & 1.30(10) \\
$q^2(0) \rho(r)$                     & 1.54(47) & 1.81(20) & 2.41(58) \\
\hline 
\end{tabular}
\\
\end{center}
\caption{Screening masses in GeV from fits to exponential decays of 
  	the various correlators for several cooling steps. 
The fit ranges are indicated by the lines in the corresponding  figures.  
} 
\vspace{4mm}
\end{table}
%%%%%%%%%%%%%%%%%%%%%%%%%%%%%%%%%%%%%%%%%%%%%%%%%%%%%%%%%%%%%%%
We find that the correlations of the color charge density $\bar\psi\psi(x)$ and 
$|\psi^{\dagger}\psi(x)|$ with the topological 
charge density are very similar, 
both in the slopes and the absolute values. This becomes clear because the 
quark condensate can be interpreted as the absolute value of the quark density. 
However, cooling (or some other kind of smoothing) 
is inevitable to obtain nontrivial correlations between the chiral density, 
${\cal O}_1=\bar \psi \gamma_5 \psi(x)$, and the topological charge density (c). 
This can be expected since both quantities are correlated via the anomaly. 
The topological charge of a gauge field is related to the 
chiral density of the associated fermion field by
\begin{equation}
Q=\int q(x) d^4x= m\int \bar\psi\gamma_5\psi(x)  d^4x. 
\end{equation}
We have checked that this relation also holds approximately for 
the corresponding lattice observables on individual configurations.
The autocorrelation 
function of the density of the 
topological charge $<q(0)q(r)>$ (d)  should be compared to  
$<\bar\psi\gamma_5\psi(0)q(r)>$ (c). 
If the classical t'Hooft instanton with size $\rho$ 
is considered,  the topological charge density 
is 
\begin{equation}
q(x) \propto \rho^4(x^2+\rho^2)^{-4} \quad .
\end{equation}
On the other hand the corresponding density of fermionic quantities \cite{SHU88}
\begin{equation}
\bar \psi \psi(x)  \propto \bar \psi\gamma_5\psi(x)  \propto \rho^2(x^2+\rho^2)^{-3}
\end{equation}
is broader. This behavior is reflected in (c) and (d) 
where the corresponding correlators are compared. 
The local relation $ q(x)=m\, \bar\psi\gamma_5\psi(x)$
does not hold as one might naively expect, the fermionic 
distributions are of longer range than the instanton profiles. 
We performed check runs with 400 fermionic updates but did not 
find significant changes in the correlators. 
%14%%%%%%%%%%%%%%%%%%%%%%%%%%%%%%%%%%%%%%%%%%%%%%%%%%%%%%%%%%%%%%
\begin{figure*}
\begin{tabular}{c}
\epsfxsize=13.5cm\epsffile{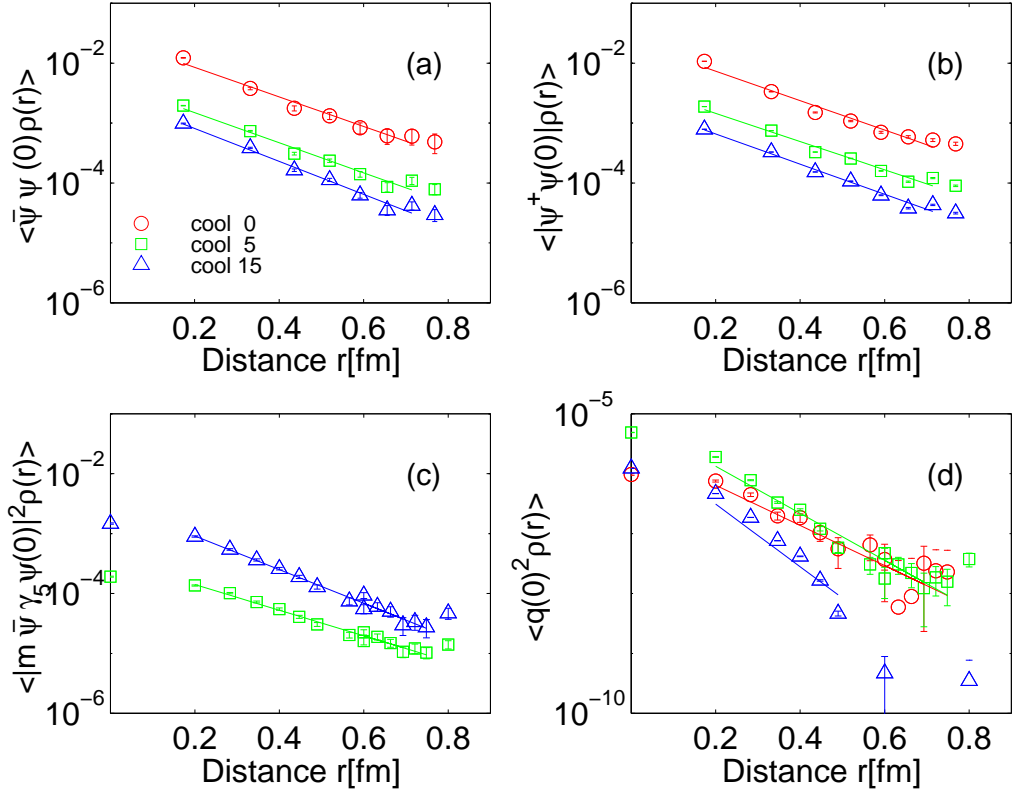} \\
\end{tabular}
\caption{
Correlation functions of the monopole  density 
with the quark condensate (a), quark charge (b)  and 
chiral density (c), and with the topological charge density (d).  
In the plots (c) and (d) some points cannot be resolved or 
error bars below zero are not printed.  }
\label{corrmon}
\end{figure*}
%%%%%%%%%%%%%%%%%%%%%%%%%%%%%%%%%%%%%%%%%%%%%%%%%%%%%%%%%%%%%%%

In Fig.~\ref{corrmon} we show the analogous situation to Fig. 1 with 
${\cal O}_2$ being the monopole density. Screening masses from  
corresponding fits to exponentials  are found in Table 1 favoring again an exponential decay of the correlations. 
As in the case with the topological charge correlators, the monopole 
correlators are nearly cooling-independent for the 
local chiral condensate (a) and the modulus of the 
quark charge density (b).
This expresses the strong 
correlations between the monopole and the topological charge densities 
themselves \cite{wir}. 
The correlations  of the chiral density with the 
monopole density (c)
deserve cooling to resolve a signal. 
We again compare the topological 
charge - monopole density correlators (d) to the chiral density  
results (c). The behavior of their widths is compatible with 
Eqs. (12) and (13) described 
above to discuss Fig.~1 (c) and (d). 

We now turn to a direct visualization of  fermionic densities   and
topological quantities on individual gauge fields rather than 
performing gauge averages. We pursue this in the following to 
get insight into the local interplay of topology with   
the sea-quark distribution.  
In a series of papers we found that at 
clusters of topological charge density, which are identified with
instantons, there are monopole trajectories looping around in almost all 
cases  for SU(2) and SU(3) gauge theories,
with improved action, in the presence of dynamical quarks, even across the 
deconfinement phase transition \cite{wir}. 
By analyzing dozens of gluon and quark field configurations we obtained the
following results.
The topological charge is hidden in  quantum fluctuations and
becomes visible by cooling of the gauge fields. For
0 cooling steps no structure can be seen in $q(x)$, the fermionic 
observables 
and  the monopole currents, which does not mean the absence of correlations
between them. After a few  cooling steps clusters of
 nonzero topological charge density and quark  fields  are resolved.
For more cooling steps both topological charge and
quark fields  begin to die out and eventually vanish.

\begin{figure}
\begin{tabular}{ccc}
(a) & (b)  & (c) \\
\epsfxsize=4.0cm\epsffile{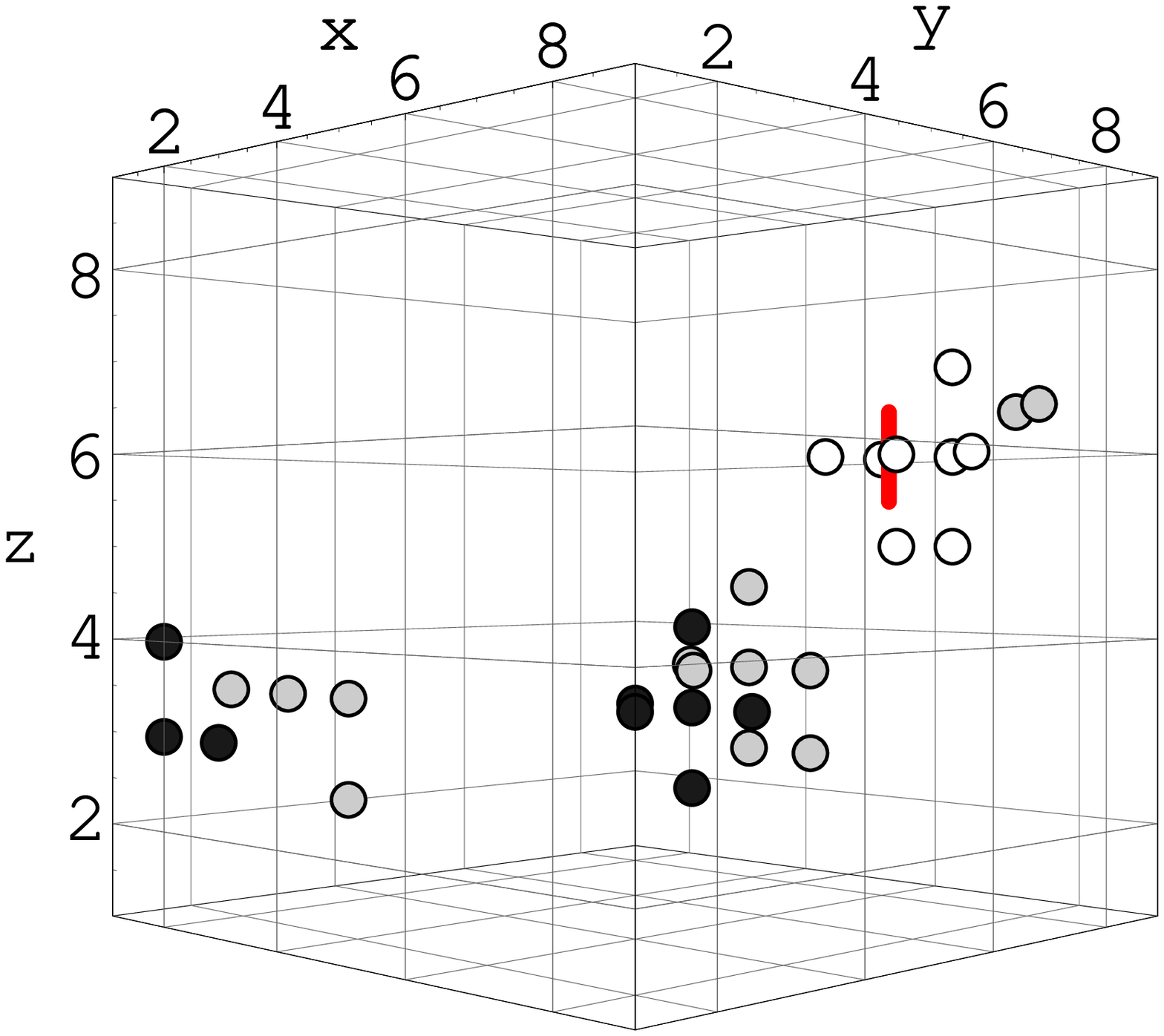 } & 
\epsfxsize=4.0cm\epsffile{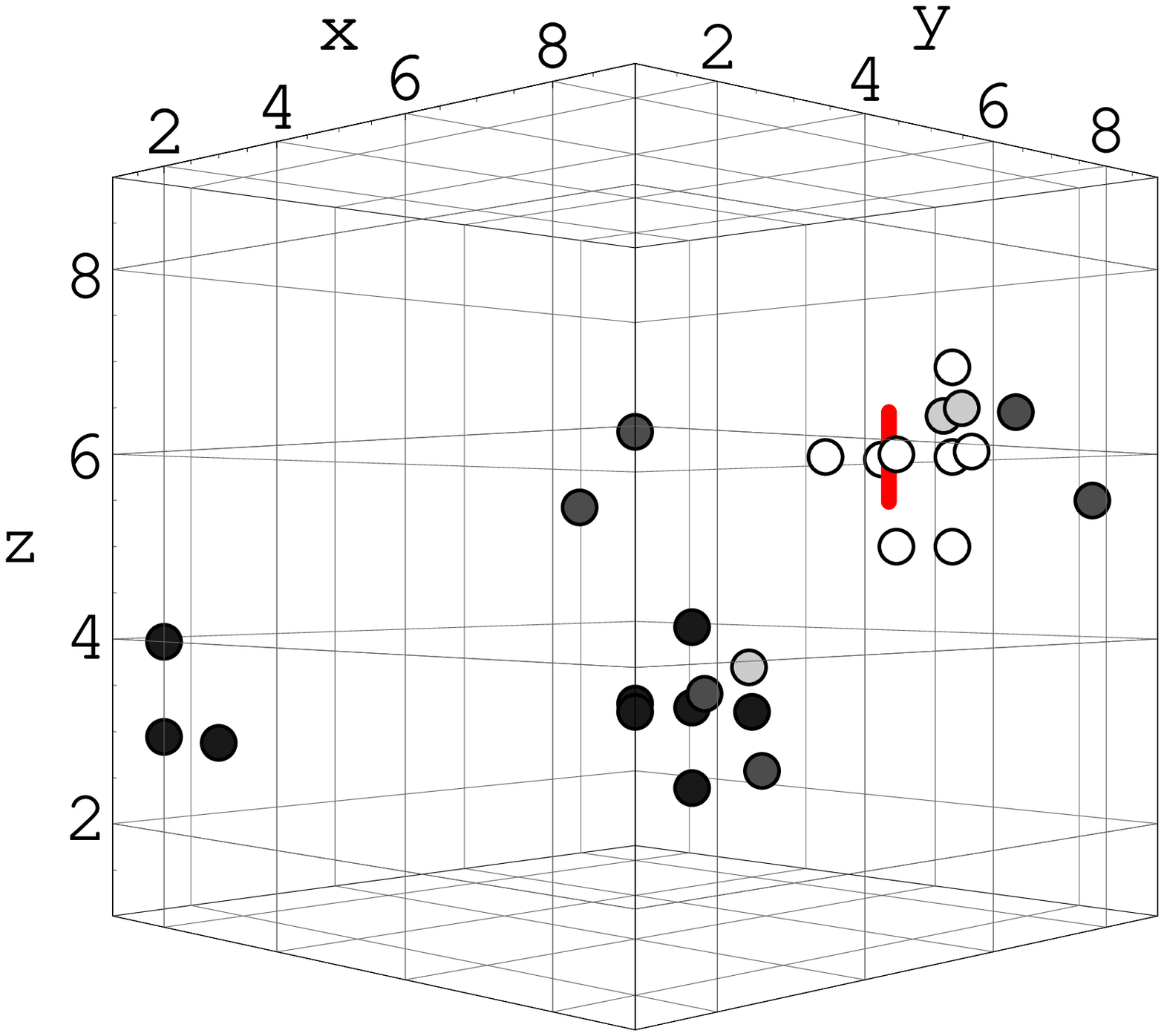 } &
\epsfxsize=4.0cm\epsffile{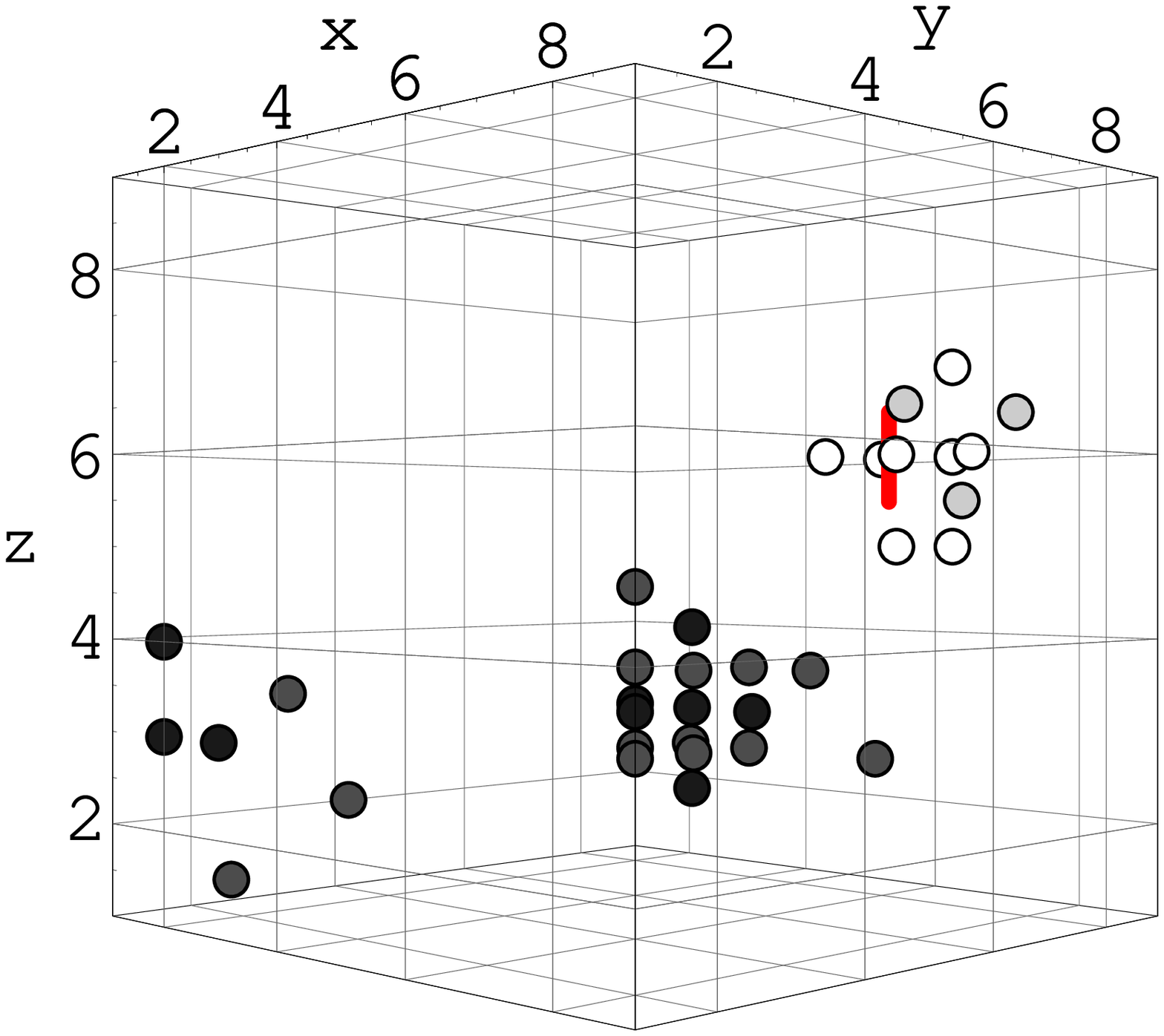 } \\
\end{tabular}
\caption{
Aspects of the fermionic field for a fixed time slice in a spatial 
volume of $(1.6 \,\, {\rm fm})^3$ for a gauge field 
with an instanton-antiinstanton pair drawn by white and black dots. 
(a) 
$\bar\psi\psi(x) > 0$ (light grey dots) within the instantons. 
(b) 
$\psi^{\dag}\psi(x)$ alternates in sign (light and dark grey dots) in a single instanton.  
(c)  $\bar \psi\gamma_5\psi(x)$ is positive/negative in instanton/antiinstanton.} 
\label{hist}
\end{figure}

In Fig.~\ref{hist} a  typical topologically nontrivial 
configuration, consisting of an instanton and an antiinstanton, 
from SU(3) theory
with dynamical quarks on the $8^{3} \times 4$ lattice 
in the confinement phase is shown after 15 cooling steps 
for fixed time slices.
%%20%%%%%%%%%%%%%%%%%%%%%%%%%%%%%%%%%%%%%%%%17 
%%%%%%%%%%%%%%%%%%%%%%%%%%%%%%%%%%%%%%%%%%
We display the positive/negative topological charge density  by  
white/black dots if the absolute value exceeds certain minimal fluctuations. 
Monopole currents are defined in the maximum Abelian projection 
and only one type is shown  by lines. 
The left 3D plot includes the local chiral condensate $\bar\psi\psi(x)$, 
indicated by light grey  dots whenever 
a certain threshold is exceeded. One clearly sees that both the instanton and 
antiinstanton are surrounded 
by a cloud of $\bar\psi\psi(x) > 0$ \cite{HANDS}. The middle 
3D plot exhibits the situation for the quark charge density 
$\psi^{\dag}\psi(x)$ indicated by light and dark grey dots depending on the 
sign of the net color charge excess. One observes that $\psi^{\dag}\psi(x)$ 
alternates in sign already in one instanton implying trivial 
correlations $<\psi^{\dag}\psi(x) q(y)>=0$ (not shown in Fig. 1). 
The right 3D plot displays the chiral density $\bar \psi \gamma_5 \psi(x)$ 
again indicated by light and dark grey dots. One nicely sees that the positive instanton 
is always surrounded by a lump with $\bar \psi \gamma_5 \psi(x)>0$ and vice versa.

\section{Discussion}

Combining the  finding of  Fig.~1 and 2 showing that the correlation 
functions between fermionic 
and topological  quantities  
are not very sensitive to cooling together with  the 
3D images in Fig.~3, we conclude that instantons go hand in hand with clusters of 
$\bar \psi\Gamma \psi (x)\neq 0$ , $\Gamma=1,\gamma_4,\gamma_5$, 
also in the uncooled QCD vacuum.  
In summary, our calculations of correlation functions
between topological densities and the fermionic observables 
exhibit exponential behavior. 
Results for the condensate and the modulus of the quark charge correlators are almost 
identical, as expected, since the quark condensate 
reflects the absolute value of the quark charge density. These 
correlation functions show little cooling dependence. 

The correlations unambiguously demonstrate that not only the local 
chiral condensate $\bar\psi\psi(x)$ but also the quark charge 
$\psi^{\dagger}\psi(x)$
and chiral density $\bar \psi \gamma_5 \psi(x)$     
take non-vanishing values predominantly in the regions  of
instantons and monopole loops.
Note that for the chiral density this behavior is expected due to the 
anomaly. 

Visualization exhibited that the distribution of 
sea-quarks is drastically enhanced around centers of nontrivial 
topology (instantons, monopoles) in Euclidian space-time. 
Our investigations of full QCD with dynamical quarks rely on lattices of 
spatial volume of $(1.6 \,\, {\rm fm})^3$ with a relatively large lattice 
constant of $0.2$ fm. The correlation functions were also computed 
in the off-axes directions and show little anisotropy effects. 
This restauration of 
rotational invariance is usually taken as a sign of set-in of asymptotic 
scaling \cite{lang}. Nevertheless, it must   
be emphasized that our results represent the situation on a finite lattice with 
finite quark mass without the extrapolation  to the thermodynamic and chiral limit. 
However, since all such correlators turned out rather independent 
of the gauge group and choice of the action etc., 
we expect that they are generic. 
We are currently studying fermionic observables along individual 
monopole trajectories, to further shed light on the puzzle how far 
monopoles are involved in  chiral symmetry breaking.

\end{document}